\documentclass[journal]{IEEEtran}
\usepackage{graphicx,amsmath,amssymb,epsfig}  
\newcommand{\Define}{\stackrel{\triangle}{=}}

\begin{document}
\title{Large MIMO Detection: A Low-Complexity Detector at High Spectral 
Efficiencies}
\author{K. Vishnu Vardhan, Saif K. Mohammed, %\IEEEmembership{Student~Member,~IEEE,}
A. Chockalingam, and %~\IEEEmembership{Senior Member,~IEEE,}
%and \\ 
B. Sundar Rajan%,~\IEEEmembership{Senior Member,~IEEE}% <-this %stops a space \\
\\
Department of ECE, Indian Institute of Science, Bangalore-560012. INDIA 
\\\vspace{2mm}{\small {\bf (Parts of this paper appeared in IEEE JSAC Special Issue on Multiuser Detection in Advanced Communication Systems and Networks, vol. 26, no. 3, pp. 473-485, April 2008, and accepted in IEEE ICC'2008)}}
%\thanks{ }
}
% note the % following the last \IEEEmembership and also the first \thanks - 
% instance, "A\textbf{} \textbf{}B" will typeset as "A B" not "AB". If you want
% "AB" then you have to do: "A\textbf{}\textbf{}B"
\markboth{}{Shell \MakeLowercase{\textit{et al.}}: Bare Demo of IEEEtran.cls for Journals}
% *** Note that you probably will NOT want to include the author's name in ***
% *** the headers of peer review papers.                                   ***

% If you want to put a publisher's ID mark on the page
% (can leave text blank if you just want to see how the
% text height on the first page will be reduced by IEEE)
%\pubid{0000--0000/00\$00.00~\copyright~2002 IEEE}

\maketitle

\begin{abstract}
We consider large MIMO systems, where by `{\em large}' we mean number
of transmit and receive antennas of the order of tens to hundreds. 
Such large MIMO systems will be of immense interest because of the very
high spectral efficiencies possible in such systems. We present a 
low-complexity detector which achieves uncoded near-exponential
diversity performance for hundreds of antennas (i.e., achieves near
SISO AWGN performance in a large MIMO fading environment) with an
average per-bit complexity of just $O(N_tN_r)$, where $N_t$ and $N_r$
denote the number of transmit and receive antennas, respectively. With
an outer turbo code, the proposed detector achieves good coded bit error
performance as well. For example, in a 600 transmit and 600 receive
antennas V-BLAST system with a high spectral efficiency of 200 bps/Hz
(using BPSK and rate-1/3 turbo code), our simulation results show that the
proposed detector performs close to within about 4.6 dB from theoretical 
capacity. We also adopt the proposed detector for the low-complexity 
decoding of high-rate non-orthogonal space-time block codes (STBC) from 
division algebras (DA). For example, we have decoded the $16\times 16$ 
full-rate non-orthogonal STBC from DA using the proposed detector and show 
that it performs close to within about 5.5 dB of the capacity using 4-QAM 
and rate-3/4 turbo code at a spectral efficiency of 24 bps/Hz. The practical 
feasibility of the proposed high-performance low-complexity detector could 
potentially trigger wide interest in the implementation of large MIMO systems. 
We also illustrate the applicability of the proposed detector in the 
low-complexity detection of large multicarrier CDMA (MC-CDMA) systems. In 
large MC-CDMA systems with hundreds of users, the proposed detector is shown 
to achieve near single-user performance at an average per-bit complexity 
linear in number of users, which is quite appealing for its use in practical 
CDMA systems.
\end{abstract}

\IEEEpeerreviewmaketitle

\section{Introduction}
\label{sec1}
\PARstart{M}{ultiple}-input multiple-output (MIMO) techniques offer 
transmit diversity and high data rates through the use of multiple 
antennas at both transmitter and receiver sides \cite{paulraj}-\cite{Fos96}. 
A key component of a MIMO system is the MIMO detector at the receiver, 
which, in practice, is often the bottleneck for the overall performance 
and complexity. 
MIMO detectors including sphere decoder and several of its variants
\cite{sphere0}-\cite{qrd} achieve near maximum likelihood (ML)
performance at the cost of high complexity. Other well known detectors
including ZF (zero forcing), MMSE (minimum mean square error), and
ZF-SIC (ZF with successive interference cancellation) detectors
\cite{tse},\cite{verdu} are attractive from a complexity view point, but
achieve relatively poor performance. For example, the ZF-SIC detector
(i.e., the well known V-BLAST detector with ordering
\cite{vblast1},\cite{vblast2}) does not achieve the full diversity in
the system. The MMSE-SIC detector has been shown to achieve optimal
performance \cite{tse}. However, these detectors are prohibitively 
complex for large number of antennas of the order of tens to hundreds. 
With small number of antennas, the high capacity potential of MIMO is
not fully exploited. A key issue with using large number of antennas, 
however, is the high detection complexities involved.

In this paper, we focus on large MIMO systems, where by `{\em large}'
we mean number of transmit and receive antennas of the order of tens to
hundreds. Such large MIMO systems will be of immense interest because
of the very high spectral efficiencies possible in such systems. For
example, in a V-BLAST system, increased number of transmit antennas
means increased data rate without bandwidth increase. However, major
bottlenecks in realizing such large MIMO systems include $i)$ physical
placement of large number of antennas in communication 
terminals\footnote{We, however, point out that there can be several large 
MIMO applications where antenna placement need not be a major issue. An 
example of such a scenario is to provide high-speed back-haul connectivity 
between base stations using large MIMO links, where large number of antennas 
can be placed at the base stations. Also, tens of antennas can be placed in 
moderately sized terminals (e.g., laptops, set top boxes) that can enable 
interesting spectrally efficient, high data rate applications like wireless 
IPTV.}, $ii)$ lack of practical low-complexity detectors for such large 
systems, and $iii)$ channel estimation issues. In this paper, we address 
the second problem in the above (i.e., low-complexity large MIMO detection). 
Specifically, we present a low-complexity detector for large MIMO systems, 
including V-BLAST as well as high-rate non-orthogonal space-time block 
codes (STBC) \cite{sundar}.

The proposed low-complexity detector has its roots in past work on 
Hopfield neural network (HNN) based algorithms for image restoration 
\cite{hnn},\cite{sun}, which are meant to handle large digital images. 
HNN based image restoration algorithms in \cite{sun} are applied to 
multiuser detection (MUD) in CDMA systems on AWGN channels in \cite{las}. 
This detector, referred to as the likelihood ascent search (LAS) detector, 
essentially searches out a sequence of bit vectors with monotonic likelihood 
ascent and converges to a fixed point in finite number of steps \cite{las}. 
The power of the LAS detector for CDMA lies in $i)$ its linear average 
per-bit complexity in number of users, and $ii)$ its ability to perform 
very close to ML detector for large number of users. Taking the cue from 
LAS detector's complexity and performance superiority in large systems, we, 
in this paper, successfully adopt the LAS detector for large MIMO systems 
and report interesting results.

We refer to the proposed detector as MF/ZF/MMSE-LAS\footnote{Throughout
the paper, whenever we write MF/ZF/MMSE-LAS, we mean MF-LAS, ZF-LAS, and
MMSE-LAS.} detector depending on the initial vector used in the algorithm;
MF-LAS detector uses the matched filter output as the initial vector, and
ZF-LAS and MMSE-LAS detectors employ ZF and MMSE outputs, respectively,
as the initial vector. Our major findings in this paper are summarized 
as follows:

\vspace{4mm}
{\bf Detection in Large V-BLAST Systems:}

\begin{itemize}
\item
In an uncoded V-BLAST system with BPSK, the proposed detector achieves
{\em near-exponential diversity} for hundreds of antennas (i.e., achieves 
near SISO AWGN performance). For example, the proposed detector nearly 
renders a $200\times200$ MIMO fading channel into 200 parallel, 
non-interfering SISO AWGN channels. The detector achieves this excellent 
performance with an average per-bit complexity of just $O(N_tN_r)$, where 
$N_t$ and $N_r$ denote the number of transmit and receive antennas, 
respectively.
\item
With an outer turbo code, the proposed detector achieves good coded bit
error performance as well. For example, in a 600 transmit and 600 receive
antennas V-BLAST system with a high spectral efficiency of 200 bps/Hz
(using BPSK and rate-1/3 turbo code), our simulation results show that
the proposed detector performs close to within about 4.6 dB from the 
theoretical capacity. We note that performance with such closeness to 
capacity has not been reported in the literature so far for such large 
number of antennas using a practical complexity detector.
\end{itemize}

\vspace{3mm}
{\bf Detection of Large Full-Rate Non-Orthogonal STBCs:}

\begin{itemize}
\item
We have adopted the proposed detector for the low-complexity decoding
of large full-rate non-orthogonal STBCs from division algebras (DA) in 
\cite{sundar}. We decode the $16\times 16$ full-rate non-orthogonal
STBC from DA (which has 256 data symbols in one STBC matrix) using the 
proposed detector and show that it performs close to within about 5.5 dB 
from capacity using 4-QAM and rate-3/4 turbo code at a spectral efficiency 
of 24 bps/Hz.
\item
We point out that because of the high complexities involved in the decoding 
of large non-orthogonal STBCs using other known detectors, the BER performance 
of such high-rate large non-orthogonal STBCs have not been reported in the 
literature so far. The very fact that we could show the simulated BER 
performance plots (both uncoded as well as turbo coded) for a $16\times 16$ 
full-rate non-orthogonal STBC with 256 complex symbols in one STBC matrix 
in itself is a clear indication of the superior low-complexity attribute of 
the proposed detector. To our knowledge, this is the first time that 
simulated BER plots and nearness to capacity results for a full-rate 
$16\times 16$ STBC from DA are reported in the literature; this became 
feasible due to the low-complexity attribute of the proposed detector.
\end{itemize}

\vspace{3mm}
{\bf Detection in Large Multicarrier CDMA Systems:}

\begin{itemize}
\item
We also illustrate the applicability of the proposed detector in the
low-complexity detection of large multicarrier CDMA (MC-CDMA) systems.
In large MC-CDMA systems with hundreds of users, the proposed detector 
is shown to achieve near single-user performance, at an average per-bit 
complexity linear in number of users, which is quite appealing for its 
use in practical CDMA systems.
\end{itemize}

The rest of the paper is organized as follows. In section \ref{sec2}, we
present the proposed LAS detector for V-BLAST systems and its complexity.
The simulated uncoded and coded BER performance of the proposed detector
for V-BLAST is presented in section \ref{sec3}. Decoding of non-orthogonal
STBCs and BER performance results are presented in section \ref{sec4}. 
The LAS detector for MC-CDMA and the corresponding BER performance results 
are presented in section \ref{sec5}. Conclusions are presented in section 
\ref{sec6}.

\section{Proposed LAS Detector for Large MIMO}
\label{sec2}
In this section, we present the proposed LAS detector for V-BLAST and its
complexity. Consider a V-BLAST system with $N_t$ transmit antennas and
$N_r$ receive antennas, $N_t\leq N_r$, where $N_t$ symbols are transmitted
from $N_t$ transmit antennas simultaneously. Let $b_j \in \{+1,-1\}$ be the
symbol\footnote{Although we present the detector for BPSK here, we have
adopted it for $M$-QAM/$M$-PAM as well.} transmitted by the $j$th transmit
antenna. Each transmitted symbol goes through the wireless channel to arrive
at each of $N_r$ receive antennas. Denote the path gain from transmit antenna
$j$ to receive antenna $k$ by $h_{kj}$. Considering a flat-fading MIMO
channel with rich scattering, the signal received at antenna $k$,
denoted by $y_k$, is given by
\begin{eqnarray}
y_k & = & \sum_{j=1}^{N_t} h_{kj}b_j + n_k.
\end{eqnarray}
The $\{h_{kj}\}$, $\forall k \in \{1,2,\cdots,N_r\}$,
$\forall j\in \{1,2,\cdots,N_t\}$, are assumed to be i.i.d. complex Gaussian
r.v's (i.e., fade amplitudes are Rayleigh distributed) with zero mean and
$E\big[\big(h_{kj}^{I}\big)^2\big] =E\big[\big(h_{kj}^{Q}\big)^2\big]=0.5$,
where $h_{kj}^{I}$ and $h_{kj}^{Q}$ are the real and imaginary parts of
$h_{kj}$. The noise sample at the $k$th receive antenna, $n_k$, is assumed
to be complex Gaussian with zero mean, and $\{n_k\}$, $k=1,2,\cdots,N_r$,
are assumed to be independent with
$E[n_k^2]=N_0=\frac{N_tE_s}{\gamma}$, where $E_s$ is the average
energy of the transmitted symbols, and $\gamma$ is the average received SNR
per receive antenna \cite{jafarkhani}. Collecting the received signals from
all receive antennas, we write\footnote{We adopt the following notation:
Vectors are denoted by boldface lowercase letters,
and matrices are denoted by boldface uppercase letters. $[.]^T$, $[.]^*$, and
$[.]^H$ denote transpose, conjugate, and conjugate transpose operations,
respectively.  $\Re(.)$ and $\Im(.)$ denote the real and imaginary
parts of the complex argument.}
\begin{eqnarray}
{\bf y} & = & {\bf H}{\bf b} + {\bf n},
\end{eqnarray}
where ${\bf y}=\left[\,y_1\,\,y_2\,\,\cdots\,\,y_{N_r}\,\right]^T$
is the $N_r$-length received signal vector,
${\bf b} = \left[\,b_1 \,\, b_2 \,\, \cdots \,\, b_{N_t}\,\right]^T$
is the $N_t$-length transmitted bit vector, {\bf H} denotes the
$N_r\times N_t$ channel matrix with channel coefficients $\{h_{kj}\}$, and
${\bf n} = \left[\,n_1 \,\, n_2 \,\, \cdots \,\, n_{N_r}\,\right]^T$
is the $N_r$-length noise vector. {\bf H} is assumed to be known perfectly
at the receiver but not at the transmitter.

\subsection{Proposed LAS Algorithm }
\label{sec3aa}
The proposed LAS algorithm essentially searches
out a sequence of bit vectors until a fixed point is reached; this sequence
is decided based on an update rule. In the V-BLAST system considered, for
ML detection \cite{verdu}, the most likely $\bf b$ is taken as that ${\bf b}$
which maximizes
\begin{eqnarray}
\Lambda({\bf b}) & = & {\bf b}^T{\bf H}^H{\bf y} + {\bf b}^T\left({\bf H}^H{\bf y}\right)^* - {\bf b}^T{\bf H}^H{\bf H}{\bf b}.
\label{lamda1w}
\end{eqnarray}
The likelihood function in (\ref{lamda1w}) can be written as
\begin{eqnarray}
\Lambda({\bf b}) & = & {\bf b}^T{\bf y}_{eff} - {\bf b}^T{\bf H}_{eff}{\bf b},\label{3eq5w}
\end{eqnarray}
where
\begin{eqnarray}
{\bf y}_{eff} &= &\,{\bf H}^H{\bf y} + \left({\bf H}^H{\bf y}\right)^*,\\
\label{3eq5aw}
{\bf H}_{eff} & = & {\bf H}^H{\bf H}.
\label{3eq5bw}
\end{eqnarray}

{\em Update Criterion in the Search Procedure:}
Let ${\bf b}(n)$ denote the bit vector tested by the LAS algorithm in
the search step $n$. The starting vector ${\bf b}(0)$ can be the output
vector from any known detector. When the output vector of the MF detector
is taken as the ${\bf b}(0)$, we call the resulting LAS detector as the
MF-LAS detector. We define ZF-LAS and MMSE-LAS detectors likewise. Given
${\bf b}(n)$, the algorithm obtains ${\bf b}(n+1)$ through an update
rule until a fixed point is reached. The update is made in such a way that
the change in likelihood from step $n$ to $n+1$, denoted by
$\Delta\Lambda\left({\bf b}(n)\right)$, is positive, i.e.,
\begin{eqnarray}
\Delta\Lambda\left({\bf b}(n)\right) & \Define &
\Lambda\left({\bf b}(n+1)  \right) - \Lambda\left({\bf b}(n)  \right) \,\,
\geq \,\, 0.
\label{deltaw}
\end{eqnarray}
An expression for the above change in likelihood can be obtained
in terms of the gradient of the likelihood function as follows.
Let ${\bf g}(n)$ denote the gradient of the likelihood
function evaluated at ${\bf b}(n)$, i.e.,
\begin{eqnarray}
{\bf g}(n) & \Define & \frac{\partial\left(\Lambda({\bf b}(n))\right)}{\partial\left({{\bf b}(n)}\right)} \,\,\, = \,\,\,
{\bf y}_{eff} - {\bf H}_{real}{\bf b}(n),
\label{3eq9w}
\end{eqnarray}
where
\begin{eqnarray}
{\bf H}_{real} & = & {\bf H}_{eff} + \left({\bf H}_{eff}\right)^*
\,\, = \,\, 2 \, \Re\left({{\bf H}_{eff}}\right).
\label{3eq10w}
\end{eqnarray}
Using (\ref{3eq5w}) in (\ref{deltaw}), we can write
\begin{eqnarray}
\Delta\Lambda\left({\bf b}(n)\right) & = & {\bf b}^T(n+1){\bf y}_{eff} - {\bf b}^T(n+1){\bf H}_{eff}{\bf b}(n+1) \nonumber \\
& & 
- \,\big({\bf b}^T(n){\bf y}_{eff} - {\bf b}^T(n){\bf H}_{eff}{\bf b}(n)\big) \nonumber \\
& = & \left({\bf b}^T(n+1) - {\bf b}^T(n)\right)\left({\bf y}_{eff} - {\bf H}_{real}{\bf b}(n)\right) \nonumber \\
& & - \, \left({\bf b}^T(n+1) - {\bf b}^T(n)\right)\left(- {\bf H}_{real}{\bf b}(n)\right) \nonumber \\
& & \hspace{-16mm} - \, {\bf b}^T(n+1){\bf H}_{eff}{\bf b}(n+1) + {\bf b}^T(n){\bf H}_{eff}{\bf b}(n).
\label{dxw}
\end{eqnarray}
Now, defining
\begin{eqnarray}
\Delta {\bf b}(n) & \Define & {\bf b}(n+1) - {\bf b}(n),
\label{3eq7w}
\end{eqnarray}
and $i)$ observing that
${\bf b}^T(n){\bf H}_{real}{\bf b}(n) = 2{\bf b}^T(n){\bf H}_{eff}{\bf b}(n)$, 
$ii)$
adding \& subtracting the term $\frac{1}{2}{\bf b}^T(n){\bf H}_{real}{\bf b}(n+1)$ to the RHS of (\ref{dxw}), and $iii)$ further observing that
${\bf b}^T(n){\bf H}_{real}{\bf b}(n+1) = {\bf b}^T(n+1){\bf H}_{real}{\bf b}(n)$,
we can simplify (\ref{dxw}) as
\begin{eqnarray}
\Delta\Lambda\left({\bf b}(n)\right) & = & \Delta {\bf b}^T(n) \left({\bf y}_{eff} - {\bf H}_{real}{\bf b}(n)\right) \nonumber \\
& & - \, \frac{1}{2}\Delta {\bf b}^T(n){\bf H}_{real}\Delta {\bf b}(n) \nonumber \\
& = & \Delta{\bf b}^T(n)\Big({\bf g}(n) + \frac{1}{2}{\bf z}(n)\Big),
\label{3eq8w}
\end{eqnarray}
where
\begin{eqnarray}
{\bf z}(n) & = & -{\bf H}_{real}\Delta {\bf b}(n).
\label{3eq11w}
\end{eqnarray}
Now, given ${\bf y}_{eff}$, ${\bf H}_{eff}$, and ${\bf b}(n)$, the
objective is to obtain ${\bf b}(n+1)$ from ${\bf b}(n)$ such that
$\Delta\Lambda({\bf b}(n))$ in (\ref{3eq8w}) is positive. Potentially
any one or several bits in ${\bf b}(n)$ can be flipped (i.e., changed
from +1 to -1 or vice versa) to get ${\bf b}(n+1)$. We refer to the set
of bits to be checked for possible flip in a step as a {\em check
candidate set}. Let $L(n) \subseteq \{1,2,\cdots,N_t\}$ denote the
check candidate set at step $n$. With the above definitions,
it can be seen that the likelihood change at step $n$, given by
(\ref{3eq8w}), can be written as
\begin{eqnarray}
\hspace{-6mm}
\Delta\Lambda({\bf b}(n)) & \hspace{-2.5mm} = \hspace{-2.5mm} & \hspace{-2mm} \sum_{j\in L(n)} \hspace{-2mm} 
\Big(b_j(n+1)-b_j(n)\Big)
\Big [g_j(n) + \frac{1}{2}z_j(n) \Big],
\label{3eq12w}
\end{eqnarray}
where $b_j(n)$, $g_j(n)$, and $z_j(n)$ are the $j$th elements of the
vectors ${\bf b}(n)$, ${\bf g}(n)$, and ${\bf z}(n)$, respectively.
As shown in \cite{las} for synchronous CDMA on AWGN, the
following update rule can be easily shown to achieve monotonic likelihood
ascent (i.e., $\Delta\Lambda({\bf b}(n)) > 0$ if there is at least one bit
flip) in the V-BLAST system as well.

{\em LAS Update Algorithm:}
Given $L(n) \subseteq \{1,2,\cdots,N_t\}, \forall n \geq 0 $ and an initial
bit vector ${\bf b}(0) \in \{-1,+1\}^{N_t}$, bits in ${\bf b}(n)$ are updated
as per the following update rule:
\begin{equation}
b_j(n+1) = \left\{
\begin{array}{ll}
+1, &  \mbox{if} \,\,\, j \in L(n), \,\, b_j(n) = -1 \\
& \mbox{and} \,\,\, g_j(n) > t_j(n),  \\ 
-1, &  \mbox{if} \,\,\, j \in L(n), \,\, b_j(n) = +1 \\ 
& \mbox{and} \,\,\, g_j(n) < -t_j(n), \\
b_j(n), & \mbox{otherwise},
\end{array}\right.
\label{3eq18w}
\end{equation}
where $t_j(n)$ is a threshold for the $j$th bit in the $n$th step
is taken to be
\begin{eqnarray}
t_j(n) & = & \sum_{i\in L(n)} \big |({\bf H}_{real})_{j,i} \big |, \,\,\, \forall j \in L(n),
\label{3eq12aw}
\end{eqnarray}
where $({\bf H}_{real})_{j,i}$ is the element in the $j$th row and $i$th
column of the matrix ${\bf H}_{real}$. 

It is noted that different choices can be made to specify the sequence
of $L(n), \forall n \geq 0$. One of the simplest sequences correspond to
checking one bit in each step for a possible flip, which is termed as a
sequential LAS (SLAS) algorithm with constant threshold,
$t_j = \big |\left({\bf H}_{real}\right)_{j,j}\big |.$
The sequence of $L(n)$ in SLAS can be such that the indices of bits
checked in successive steps are chosen circularly or randomly.
Checking of multiple bits for possible flip is also possible.
Let $L_f(n) \subseteq L(n)$ denote the set of indices of the bits
flipped according to the update rule in (\ref{3eq18w}) at step $n$.
Then the updated bit vector ${\bf b}(n+1)$ can be written as
\begin{eqnarray}
{\bf b}(n+1) & = & {\bf b}(n) - 2 \sum_{i \in L_f(n)} b_i(n){\bf e}_i,
\label{3eq20w}
\end{eqnarray}
where ${\bf e}_i$ is the $i$th coordinate vector.
Using (\ref{3eq20w}) in (\ref{3eq9w}), the gradient vector for the
next step can be obtained as
\begin{eqnarray}
{\bf g}(n+1) & = & {\bf y}_{eff}-{\bf H}_{real}{\bf b}(n+1) \nonumber \\
&=&{\bf g}(n)+2 \sum_{i \in L_f(n)} b_i(n)\left({\bf H}_{real}\right)_i \,\, ,
\label{3eq21w}
\end{eqnarray}
where $\left({\bf H}_{real}\right)_i$ denotes the $i$th column of the matrix
${\bf H}_{real}$.
The LAS algorithm keeps updating the bits in each step based on the update
rule given in (\ref{3eq18w}) until ${\bf b}(n) = {\bf b}_{fp}, \forall
n \geq n_{fp}$ for some $n_{fp} \geq 0$, in which case ${\bf b}_{fp}$ is
a fixed point, and it is taken as the detected bit vector and the algorithm
terminates.

\subsection{Complexity of the Proposed Detector for V-BLAST}
\label{vb_comp}
In terms of complexity, given an initial vector, the 
the LAS operation part
alone has an average per-bit complexity of $O(N_tN_r)$. This can be
explained as follows. The complexity involved in the LAS operation is
due to three components: $i)$ initial computation of ${\bf g}(0)$ in
(\ref{3eq9w}), $ii)$ update of ${\bf g}(n)$ in each step as per
(\ref{3eq21w}), and $iii)$ the average number of steps required to reach
a fixed point. Computation of ${\bf g}(0)$ requires the computation of
${\bf H}^H{\bf H}$ for each MIMO fading channel realization $\big($see
Eqns. (\ref{3eq9w}), (\ref{3eq10w}), and (\ref{3eq5bw})$\big )$, which
requires a per-bit complexity of order $O(N_tN_r)$. Update of ${\bf g}(n)$
in the $n$th step as per (\ref{3eq21w}) using sequential LAS requires a
complexity of $O(N_t)$, and hence a constant per-bit complexity.
Regarding the complexity component $iii)$, 
we obtained the average number of steps required to reach a fixed
point for sequential LAS through simulations. We observed that the
average number of steps required is linear in $N_t$, i.e., constant
per-bit complexity where the constant $c$ depends on SNR, $N_t$, $N_r$,
and the initial vector (see Fig. \ref{fig_bfr}). Putting the complexities 
of $i)$, $ii)$, and $iii)$ in the above together, we see that the average 
per-bit complexity of LAS operation alone is $O(N_tN_r)$. In addition to 
the above, the initial vector generation also contributes to the overall 
complexity. The average per-bit complexity of generating initial vectors 
using MF, ZF, and MMSE are $O(N_r)$, $O(N_tN_r)$, and $O(N_tN_r)$, 
respectively. The higher complexity of ZF and MMSE compared to MF is 
because of the need to perform matrix inversion operation in ZF/MMSE.
Again, putting the complexities of the LAS part and the initial
vector generation part together, we see that the overall average per-bit
complexity of the proposed MF/ZF/MMSE-LAS detector is $O(N_tN_r)$.
This complexity is an order superior compared to the well known ZF-SIC 
detector with ordering\footnote{Henceforth, we use the term `ZF-SIC' to 
always refer `ZF-SIC with ordering'.}, whose per-bit complexity is 
$O(N_t^2N_r)$.

\section{LAS Detector Performance in V-BLAST}
\label{sec3}
In this section, we present the uncoded/coded BER performance of the 
proposed LAS detector in V-BLAST obtained through simulations, and compare 
with those of other detectors. The LAS algorithm used is the sequential LAS 
with circular checking of bits starting from the first antenna bit. We also 
quantify how far is the proposed detector's turbo coded BER performance away 
from the theoretical capacity. The SNRs in all the BER performance figures 
are the average received SNR per received antenna, $\gamma$, defined in 
Sec. \ref{sec2} \cite{jafarkhani}.

\subsection{Uncoded BER Performance}
\begin{figure}
%\centering
\hspace{-4mm}
\includegraphics[width=3.525in]{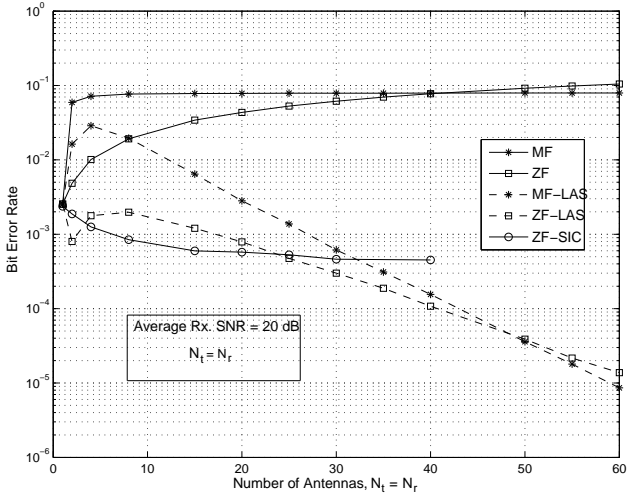}
\caption{Uncoded BER performance of MF/ZF-LAS detectors as a function of
number of transmit/receive antennas ($N_t=N_r$) for V-BLAST at an average
received SNR = 20 dB. BPSK, $N_t$ bps/Hz spectral efficiency.  }
\vspace{-2mm}
\label{fig_vb2}
\end{figure}

{\em MF/ZF-LAS performs increasingly better than ZF-SIC for increasing
$N_t=N_r$}:
In Fig. \ref{fig_vb2}, we plot the uncoded BER performance of the MF-LAS,
ZF-LAS and ZF-SIC detectors for V-BLAST as a function of $N_t=N_r$ at an
average received SNR of 20 dB with BPSK. The performance of the MF and ZF
detectors are also plotted for comparison. From Fig. \ref{fig_vb2}, we 
observe the following: 
\begin{itemize}
\item
The BER at $N_t=N_r=1$ is nothing but the SISO flat Rayleigh fading BER for
BPSK, given by
{\small $\frac{1}{2}\left[1-\sqrt{\frac{\gamma}{1+\gamma}}\right]$}
which is equal to $2.5\times 10^{-3}$ for $\gamma=20$ dB \cite{proakis}.
While the performance of MF and ZF degrade as $N_t=N_r$ is increased, the
performance of ZF-SIC improves for antennas up to $N_t=N_r=15$, beyond which
a flooring effect occurs. This improvement is likely due to the potential
diversity in the ordering (selection) in ZF-SIC, whereas the flooring for
$N_t>15$ is likely due to interference being large beyond the cancellation
ability of the ZF-SIC.
\item
The behavior of MF-LAS and ZF-LAS for increasing $N_t=N_r$  are interesting.
Starting with the MF output as the initial vector, the MF-LAS always achieves
better performance than MF. More interestingly, this improved performance
of MF-LAS compared to that of MF increases remarkably as $N_t=N_r$ increases.
For example, for $N_t=N_r=15$, the performance improves by an order in BER
(i.e., $7.5\times 10^{-2}$ BER for MF versus $7\times 10^{-3}$ BER for MF-LAS),
whereas for $N_t=N_r=60$ the performance improves by four-orders in BER
(i.e., $8\times 10^{-2}$ BER for MF versus $9\times 10^{-6}$ BER for
MF-LAS). This is due to the large system effect in the LAS algorithm which
is able to successfully pick up much of the diversity possible in the system.
This large system performance superiority of the LAS is in line with the
observations/results reported in \cite{las} for a large CDMA
system $\big($large number of antennas in our case, whereas it was large
number of users in \cite{las}$\big)$.
\item
While the ZF-LAS performs slightly better than ZF-SIC for antennas less
than 4, ZF-SIC performs better than ZF-LAS for antennas in the range 4 to
24. This is likely because, for antennas less than 4, the BER of ZF is
small enough for the LAS to clean up the ZF initial vector into an output
vector better than the ZF-SIC output vector. However, for antennas in the
range of 4 to 24, the BER of ZF gets high to an extent that the ZF-LAS is
less effective in cleaning the initial vector beyond the diversity
performance achieved by the ZF-SIC. A more interesting observation, however,
is that for antennas greater than 25, the large system effect of ZF-LAS
starts showing up. So, in the large system setting (e.g., antennas more 
than 25 in Fig. \ref{fig_vb2}), the ZF-LAS performs increasingly better 
than ZF-SIC for increasing $N_t=N_r$. We found the number of antennas at
which the cross-over between ZF-SIC and ZF-LAS occurs to be different for
different SNRs.
\item
Another observation in Fig. \ref{fig_vb2} is that for antennas greater than
50, MF-LAS performs better than ZF-LAS. This behavior can be explained by
observing the performance comparison between MF and ZF detectors given in
the same figure. For more than 50 antennas, MF performs slightly better
than ZF. It is known that ZF detector can perform worse than MF detector
under high noise/interference conditions \cite{verdu} (here high
interference due to large $N_t$). Hence, starting with a better initial
vector, MF-LAS performs better than ZF-LAS.
\end{itemize}
\begin{figure}[t]
\centering
\includegraphics[width=3.325in]{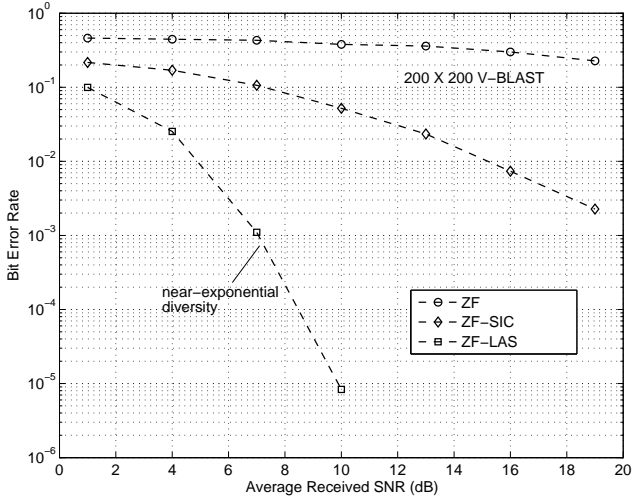}
\caption{Uncoded BER performance of ZF-LAS versus ZF-SIC as a function of
average received SNR for a $200\times 200$ V-BLAST system. BPSK, 200 bps/Hz
spectral efficiency. ZF-LAS achieves higher order diversity (near-exponential
diversity) than ZF-SIC at a much lesser complexity.}
\vspace{-2mm}
\label{fig_vb3}
\end{figure}

{\em ZF-LAS outperforms ZF-SIC in large V-BLAST systems both in complexity
\& diversity:}
In Fig. \ref{fig_vb3}, we present an interesting comparison of the
uncoded BER performance between ZF, ZF-LAS and ZF-SIC, as a function
of average SNR for a $200\times 200$ V-BLAST system. This system being
a large system, the ZF-LAS has a huge complexity advantage over ZF-SIC
as pointed out before in Sec. \ref{vb_comp}. In fact, although we have
taken the effort to show the performance of ZF-SIC at such a large number
of antennas like 200, we had to obtain these simulation points for ZF-SIC
over days of simulation time, whereas the same simulation points for ZF-LAS
were obtained in just few hours. This is due to the $O(N_t^2N_r)$ complexity
of ZF-SIC versus $O(N_tN_r)$ complexity of ZF-LAS, as pointed out in Sec.
\ref{vb_comp}. More interestingly, in addition to this significant complexity
advantage, ZF-LAS is able to achieve a much higher order of diversity (in
fact, near-exponential diversity) in BER performance compared to ZF-SIC
(which achieves only a little better than first order diversity). This is
clearly evident from the slopes of the BER curves of ZF-LAS and ZF-SIC.
{\em Note that the BER curve for ZF-LAS is almost the same as the uncoded
BER curve for BPSK on a SISO AWGN channel, given by $Q(\sqrt{\gamma})$
\cite{proakis}. This means that the proposed detector nearly renders a
$200\times 200$ MIMO fading channel into 200 parallel, non-interfering
SISO AWGN channels.}

\begin{figure}
\centering
\includegraphics[width=3.40in]{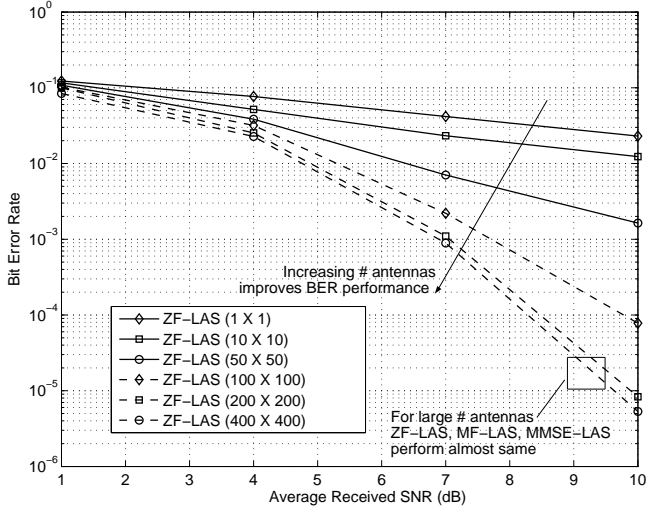}
\caption{Uncoded BER performance of ZF-LAS for V-BLAST as a function of
average received SNR for increasing values of $N_t=N_r$. BPSK, $N_t$ bps/Hz
spectral efficiency. For large number of antennas (e.g., $N_t=N_r=200,400$),
the performance of ZF-LAS, MF-LAS, and MMSE-LAS are almost the same.}
\label{fig_vb4}
\end{figure}

{\em LAS Detector's performance with hundreds of antennas:}
As pointed earlier, obtaining ZF-SIC results for more than even
50 antennas requires very long simulation run times, which is not the case
with ZF-LAS. In fact, we could easily generate BER results for up to 400
antennas for ZF-LAS, which are plotted in Fig. \ref{fig_vb4}. The key
observations in Fig. \ref{fig_vb4} are that $i)$ the
average SNR required to achieve a certain BER performance keeps
reducing for increasing number of antennas for ZF-LAS, and $ii)$
increasing the number of antennas results in increased orders
of diversity achieved (close to SISO AWGN performance for 200 and
400 antennas). We have also observed from our simulations that
for large number of antennas, the LAS algorithm converges to almost
the same near-ML performance regardless of the initial vector chosen.
For example, for the case of 200 and 400 antennas in Fig. \ref{fig_vb4},
the BER performance achieved by ZF-LAS, MF-LAS, and MMSE-LAS are
almost the same (although we have not explicitly plotted the BER curves
for MF-LAS and MMSE-LAS in Fig. \ref{fig_vb4}). So, in such large MIMO
systems setting, MF-LAS may be preferred over ZF/MMSE-LAS since
ZF/MMSE-LAS require matrix inverse operation whereas MF-LAS does not.

\begin{figure}
\centering
\includegraphics[width=3.4in]{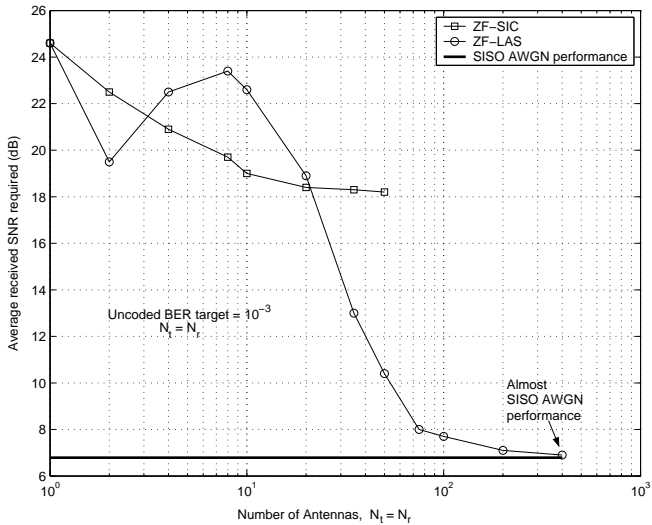}
\caption{Average received SNR required to achieve a target uncoded BER
of $10^{-3}$ in V-BLAST for increasing values of $N_t=N_r$. BPSK. ZF-LAS
versus ZF-SIC. ZF-LAS achieves near SISO AWGN performance.}
\label{fig_vb5}
\end{figure}

\begin{figure}
\centering
\includegraphics[width=3.35in]{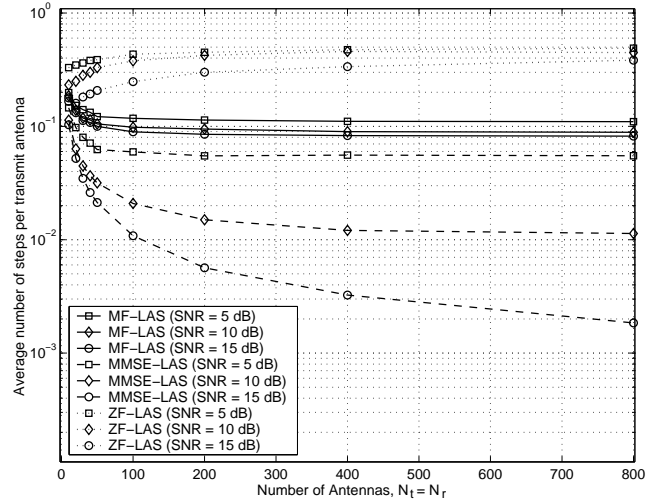}
\caption{Complexity of the LAS algorithm in terms of average
number of steps per transmit antenna till fixed point is reached
in V-BLAST as a function of $N_t=N_r$ for different SNRs and initial vectors
(MF, ZF, MMSE). BPSK. Results obtained from simulations.}
\vspace{-2mm}
\label{fig_bfr}
\end{figure}

Observation $i)$ in the above paragraph is explicitly brought out in Fig.
\ref{fig_vb5}, where we have plotted the average received SNR
required to achieve a target uncoded BER of $10^{-3}$ as a
function of $N_t=N_r$ for ZF-LAS and ZF-SIC. It can be seen that
the SNR required to achieve $10^{-3}$ with ZF-LAS significantly
reduces for increasingly large $N_t=N_r$. For example, the required
SNR reduces from about 25 dB for a SISO Rayleigh fading channel to
about 7 dB for a $400\times 400$ V-BLAST system using ZF-LAS. As we
pointed out in Fig. \ref{fig_vb4}, this $400\times 400$ system
performance is almost the same as that of a SISO AWGN channel where
the SNR required to achieve $10^{-3}$ BER is also close to 7 dB
\cite{proakis}, i.e., $20\log\left(Q^{-1}(10^{-3})\right) \approx 7$ dB.


\subsection{Turbo Coded BER Performance}
\label{turbo_perf}
In this subsection, we present the turbo coded BER performance of
the proposed LAS detector. We also quantify how far is the proposed
detector's performance away from the theoretical capacity.
For a $N_t\times N_r$ MIMO system model in Sec. \ref{sec2} with
perfect channel state information (CSI) at the receiver, the ergodic
capacity is given by \cite{tele99}
\begin{eqnarray}
C & = & E\left[\log \det\left({\bf I}_{N_r} + \left(\gamma/N_t\right){\bf H}{\bf H}^H \right)   \right],
\label{cap}
\end{eqnarray}
where ${\bf I}_{N_r}$ is the $N_r\times N_r$ identity matrix and
$\gamma$ is the average SNR per receive antenna. We have evaluated
the capacity in (\ref{cap}) for a $600 \times 600$ MIMO system through
Monte-Carlo simulations and plotted it as a function of average SNR in
Fig. \ref{fig1}. Figure \ref{fig2} shows the simulated BER performance of 
the proposed LAS detector for a $600 \times 600$ MIMO system with BPSK and 
rate-1/3 turbo code (i.e., spectral efficiency = 200 bps/Hz). Figure 
\ref{fig4} shows similar performance plots for rate-3/4 turbo code
at a spectral efficiency of 450 bps/Hz. From the capacity curve in Fig. 
\ref{fig1}, the minimum SNRs required at 200 bps/Hz and 450 bps/Hz spectral 
efficiencies are -5.4 dB and -0.8 dB, respectively. The following interesting 
observations can be made from Figs. \ref{fig2} and \ref{fig4}:

\begin{figure}
\centering
\includegraphics[width=3.25in]{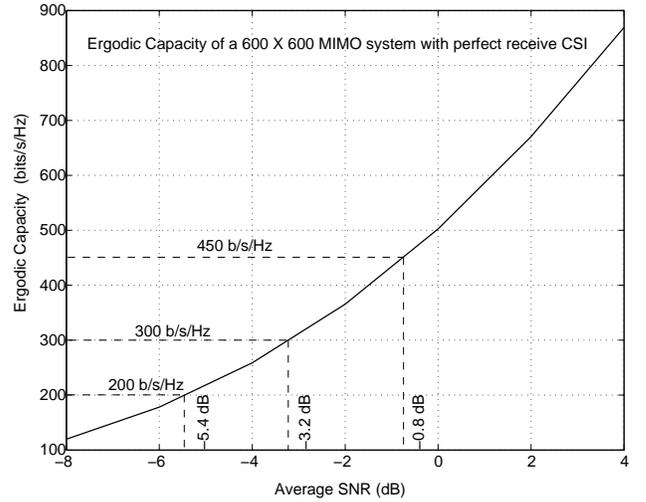}
\caption{Ergodic capacity for $600\times 600$ MIMO system with 
receive CSI.}
\label{fig1}
\end{figure}

\begin{figure}
\centering
\includegraphics[width=3.35in]{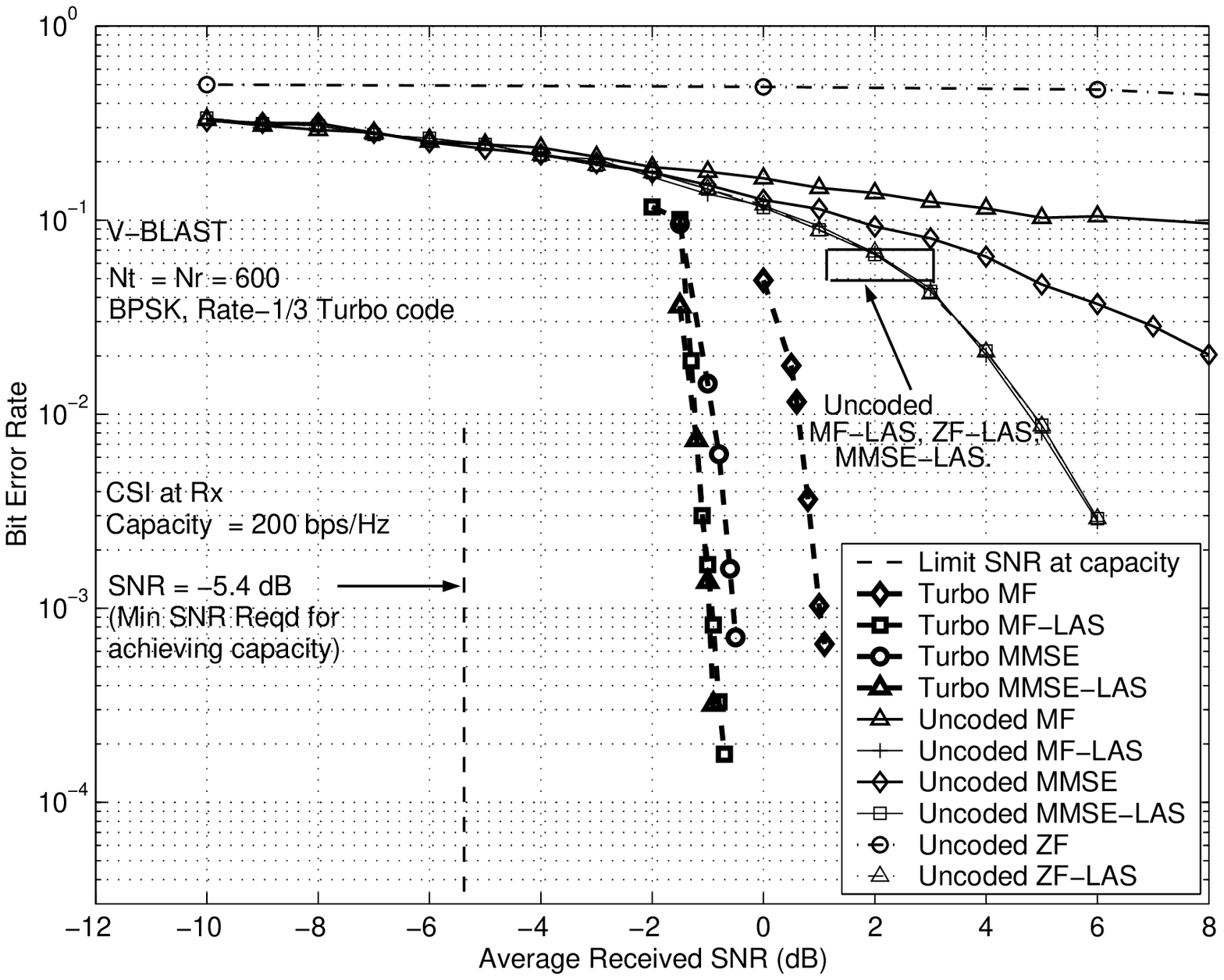}
\caption{BER performance of various detectors for rate-1/3 turbo-encoded
data using BPSK symbols in a $600\times 600$ V-BLAST system. 200 bps/Hz
spectral efficiency. Proposed MF/ZF/MMSE-LAS detectors' performance is
away from capacity by 4.6 dB.}
\label{fig2}
\end{figure}
\begin{figure}
\centering
\includegraphics[width=3.35in]{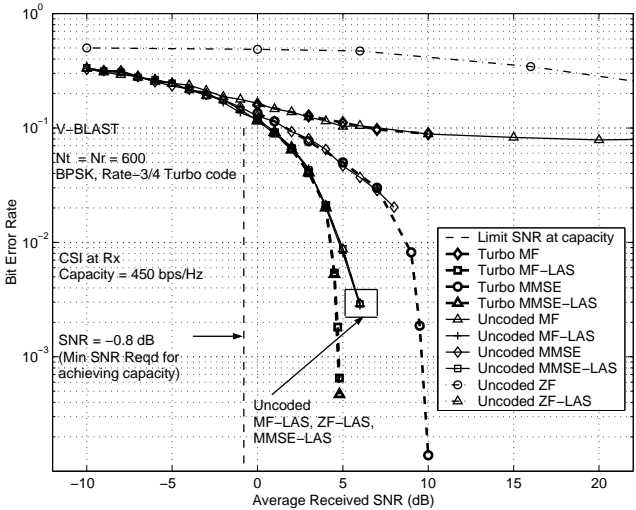}
\caption{BER performance of various detectors for rate-3/4 turbo-encoded
data using BPSK symbols in a $600 \times 600$ V-BLAST system. 450 bps/Hz
spectral efficiency. Proposed MF/ZF/MMSE-LAS detectors' performance is
away from capacity by 5.6 dB.}
\label{fig4}
\end{figure}

\begin{figure}
\centering
\includegraphics[width=3.35in]{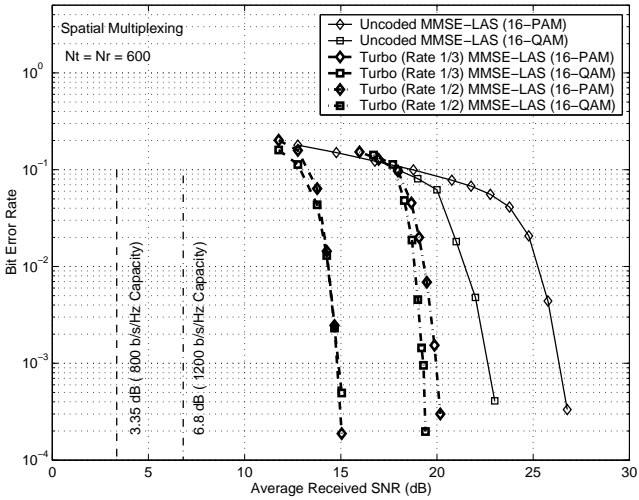}
\caption{Uncoded and coded BER performance of MMSE-LAS detector in a 
$600\times 600$ V-BLAST system for 16-PAM and 16-QAM with rate-1/2 and 
rate-1/3 turbo codes.}
\label{fig_qam}
\vspace{-3mm}
\end{figure}

\begin{itemize}
\item   In terms of uncoded BER, the performance of MF, ZF, and MMSE are
        different, with ZF and MMSE performing the worst and best,
        respectively. But the performance of MF-LAS, ZF-LAS, and MMSE-LAS
        are almost the same (near-exponential diversity performance) with
        the number of antennas being large ($N_t=N_r=600$).
\item   With a rate-1/3 turbo code (Fig. \ref{fig2}), all the LAS
        detectors considered (i.e., MF-LAS, ZF-LAS, MMSE-LAS) achieve
        almost the same performance, which is about 4.6 dB away from 
	capacity\footnote{We point out that the turbo coded BER curves 
	shown in Figs. 7 to 11 in \cite{jsac} have been plotted erroneously 
	with an SNR shift of $-10\log r$ dB, where $r$ is the turbo code 
	rate, which amounted to a pessimistic prediction of nearness to 
	capacity. Here, we have corrected those plotting errors. 
	Figures \ref{fig2}, \ref{fig4}, \ref{fig_qam} and the nearness 
	to capacity results given in Table-I in this paper are the corrected 
	ones.} (i.e., near-vertical fall of coded BER occurs 
	at about -0.8 dB). Turbo coded MF/MMSE without LAS also achieve good 
	performance in this case (i.e., less than only 2 dB away from turbo 
	coded MF/ZF/MMSE-LAS performance). This is because the uncoded BER 
	of MF and MMSE at around 0 to 2 dB SNR are small enough for the turbo 
	code to be effective. However, this is not the case with turbo coded 
	ZF without LAS. As can be seen, in  the range of SNRs shown, the
        uncoded BER of ZF without LAS is so high (close to 0.5) that the
        vertical fall of coded BER can happen only at very high SNRs,
        because of which we have not shown the performance of turbo coded
        ZF without LAS.
\end{itemize}

\begin{table}
\begin{center}
\begin{tabular}{|c|c|c|c|c|c|}
\hline
{\rule[-1mm]{0mm}{3.5mm}}
Code Rate, & Min. SNR     & \multicolumn{4}{c|}{Vertical fall of coded BER occurs at} \\ \cline{3-6}
{\rule[-1mm]{0mm}{3.5mm}}
Spect. Eff.& at capacity  & Proposed LAS & ZF   & MF   & MMSE   \\ \hline
{\rule[-1mm]{0mm}{4mm}}
Rate-1/3,  & -5.4 dB      &  -0.8 dB & high & 1.2 dB & -0.3 dB \\
200 bps/Hz &              &          &      &      &        \\ \hline
{\rule[-1mm]{0mm}{4mm}}
Rate-1/2   & -3.2 dB      &  1.5 dB  & high & high & 3 dB   \\
300 bps/Hz &              &          &      &      &        \\ \hline
{\rule[-1mm]{0mm}{4mm}}
Rate-3/4   & -0.8 dB      & 4.8 dB     & high & high & high   \\
450 bps/Hz &              &          &      &      &        \\ \hline
\end{tabular}
\end{center}
\caption{Nearness to capacity of various detectors for  $600\times600$
V-BLAST with BPSK and various turbo code rates. Proposed LAS detector
performs to within about 4.6 dB, 4.7 dB, 5.6 dB from capacity for 200,
300, and 450 bps/Hz spectral efficiencies, respectively. }
\label{tab1}
\vspace{-6mm}
\end{table}

In Table I, we summarize the performance of various detectors in terms
of their nearness to capacity in a $600\times 600$ V-BLAST system using 
BPSK, and rate-1/3, rate-1/2 and rate-3/4 turbo codes. From Table-I, it 
can be seen that there is a clear superiority of the proposed MF/ZF/MMSE-LAS 
over MF/MMSE without LAS in terms of coded BER (nearness to capacity) when 
high-rate turbo codes are used.  For example, when a rate-3/4 turbo code is 
used the MF/ZF/MMSE-LAS performs to within about 5.6 dB from capacity, 
whereas the performance of rate-3/4 turbo coded MF/MMSE without LAS are 
much farther away from capacity.

{\em Performance of $M$-PAM/$M$-QAM:}
Although the LAS algorithm in Sec. \ref{sec2} is presented assuming
BPSK, it can be adopted for $M$-ary modulation including $M$-PAM and
$M$-QAM. In the case of BPSK, the elements of the data vector take
values from $\{\pm 1\}$. $M$-PAM symbols take discrete values from
$\{A_m, 1\leq m \leq M\}$ where $A_m=(2m-1-M)$, $m=1,2,\cdots,M$, and
$M$-QAM is nothing but quadrature PAM. 
%Similar to the search over binary vectors for BPSK, likelihood ascent 
%search can be performed over $M$-ary symbol vectors. 
We have adopted the LAS algorithm for $M$-PAM/$M$-QAM
and evaluated the performance of the LAS detector for 4-PAM/4-QAM and
16-PAM/16-QAM without and with coding. In $M$-PAM/$M$-QAM also, we have
observed large system behavior of the proposed detector similar to those
presented for BPSK. As an example, in Fig. \ref{fig_qam}, we present the
uncoded and coded performance of the MMSE-LAS detector in a $600\times 600$
V-BLAST system for 16-PAM/16-QAM with rate-1/2 and rate-1/3 turbo codes at
spectral efficiencies of 1200 bps/Hz and 800 bps/Hz, respectively. It can 
be observed that the LAS detector achieves performance close to within about
13 dB from the theoretical capacity. 

{\em Effect of Channel Estimation Errors:}
As we pointed out earlier, another key issue in large MIMO systems is 
channel estimation \cite{chl_est1},\cite{chl_est2}. We have evaluated
the effect of channel estimation errors on the performance of the 
proposed detector in V-BLAST by considering an estimation error model, 
where the estimated channel matrix, ${\widehat {\bf H}}$, is taken to be
${\widehat {\bf H}} = {\bf H} + \Delta{\bf H},$ where $\Delta{\bf H}$ is 
the estimation error matrix, the entries of which are assumed to be i.i.d. 
complex Gaussian with zero mean and variance $\sigma_e^2$. Our simulation 
results showed that in a $200\times 200$ V-BLAST system with BPSK, rate-1/2 
turbo code and LAS detection, the coded BER degradation compared to perfect 
channel estimation is only 0.2 dB and 0.6 dB for channel estimation error 
variances of $1\%$ and $5\%$, respectively. The investigation of estimation 
algorithms and efficient pilot schemes for accurate channel estimation in 
large MIMO systems as such are important topics for further research. 

\vspace{-1mm}
\section{Detection of Full-Rate Non-orthogonal STBCs}
\label{sec4}
\vspace{-0mm}
V-BLAST with large number of antennas can offer high spectral efficiencies, 
but it does not provide transmit diversity. On the other hand, well known 
orthogonal STBCs have the advantages of full transmit diversity and low 
decoding complexity, but suffer from rate loss for increased number of 
transmit antennas \cite{jafarkhani},\cite{Alamouti}-\cite{tarokh3}. 
{\em Full-rate non-orthogonal STBCs from division algebras (DA)} 
\cite{sundar}, on the other hand, are attractive for achieving high 
spectral efficiencies in addition to achieving full transmit diversity, 
using large number of transmit antennas. 

Construction of full-rate non-orthogonal STBCs from DA for arbitrary number 
of transmit antennas $n$ is given by the matrix in (20.a) at the bottom of 
this page \cite{sundar}. In (20.a), $\omega_n=e^{\frac{{\bf j}2\pi}{n}}$, 
${\bf j}=\sqrt{-1}$, and 
$x_{u,v}$, $0\leq u,v \leq n-1$ are the data symbols from a QAM alphabet. 
Note that there are $n^2$ data symbols in one STBC matrix. When 
$\delta=e^{\sqrt{5}\,{\bf j}}$ and $t=e^{{\bf j}}$, the STBC in (20.a) 
achieves full transmit diversity (under ML decoding) as well as 
information-losslessness \cite{sundar}. When $\delta=t=1$, the code ceases 
to be of full-diversity (FD), but continues to be information-lossless (ILL) 
\cite{hassibi2}. 
\thanks{
\line(1,0){505}
\[
\label{eqn}
\hspace{1.3cm}
\left[
\begin{array}{ccccc}
\sum_{i=0}^{n-1}x_{0,i}\,t^i & \delta\sum_{i=0}^{n-1}x_{n-1,i}\,\omega_n^i\,t^i & \delta\sum_{i=0}^{n-1}x_{n-2,i}\,\omega_n^{2i}\,t^i & \cdots & \delta\sum_{i=0}^{n-1}x_{1,i}\,\omega_n^{(n-1)i}\,t^i \\
\sum_{i=0}^{n-1}x_{1,i}\,t^i & \sum_{i=0}^{n-1}x_{0,i}\,\omega_n^i\,t^i & \delta\sum_{i=0}^{n-1}x_{n-1,i}\,\omega_n^{2i}\,t^i & \cdots & \delta\sum_{i=0}^{n-1}x_{2,i}\,\omega_n^{(n-1)i}\,t^i \\
\sum_{i=0}^{n-1}x_{2,i}\,t^i & \sum_{i=0}^{n-1}x_{1,i}\,\omega_n^i\,t^i & \sum_{i=0}^{n-1}x_{0,i}\,\omega_n^{2i}\,t^i & \cdots & \delta\sum_{i=0}^{n-1}x_{3,i}\,\omega_n^{(n-1)i}\,t^i \\
\vdots & \vdots & \vdots & \vdots & \vdots \\
\sum_{i=0}^{n-1}x_{n-2,i}\,t^i & \sum_{i=0}^{n-1}x_{n-3,i}\,\omega_n^i\,t^i & \sum_{i=0}^{n-1}x_{n-4,i}\,\omega_n^{2i}\,t^i & \cdots & \delta \sum_{i=0}^{n-1}x_{n-1,i}\,\omega_n^{(n-1)i}t^i \\
\sum_{i=0}^{n-1}x_{n-1,i}\,t^i & \sum_{i=0}^{n-1}x_{n-2,i}\,\omega_n^i\,t^i & \sum_{i=0}^{n-1}x_{n-3,i}\,\omega_n^{2i}\,t^i & \cdots & \sum_{i=0}^{n-1}x_{0,i}\,\omega_n^{(n-1)i}\,t^i
\end{array}
\right]. \hspace{10mm} (\mbox{20.a})
\] 
}
High spectral efficiencies with large $n$ can be achieved 
using this code construction. For example, with $n=16$ transmit antennas, 
the $16\times 16$ STBC from (20.a) with 16-QAM and rate-3/4 turbo code 
achieves a spectral efficiency of 48 bps/Hz. This high spectral efficiency 
is achieved along with the full-diversity of order $nN_r$. 

However, since these STBCs are non-orthogonal, ML detection gets increasingly 
impractical for large number of transmit antennas, $n$. Consequently, a key 
challenge in realizing the benefits of these full-rate non-orthogonal STBCs 
in practice is that of achieving near-ML performance for large number of 
transmit antennas at low detection complexities. Here, we show that near-ML 
detection of large MIMO signals originating from several tens of antennas 
using full-rate non-orthogonal STBCs is possible at practically affordable 
low complexities (using the proposed LAS detector), which is a significant 
new advancement that has not been reported in the MIMO detection literature 
so far.

\subsection{Uncoded BER Results for Large STBCs from DA}
\begin{figure}
\centering
\includegraphics[width=3.35in]{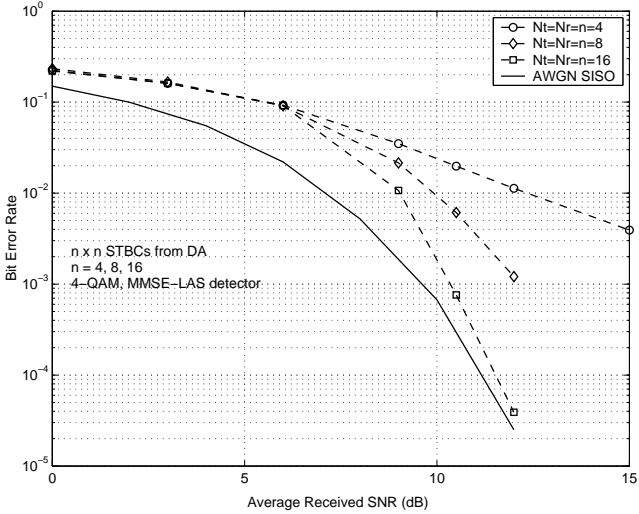}
\caption{Uncoded BER performance of the proposed LAS detector in
decoding $n\times n$ full-rate non-orthogonal STBCs from DA for
$n=4,8,16$. MMSE initial vector, 4-QAM, $N_t=N_r=n$. $16\times 16$
STBC with 256 complex symbols in each STBC matrix achieves close to 
SISO AWGN performance. }
\label{fig_n1}
\end{figure}

We have adopted the proposed LAS detector for the decoding of full-rate 
non-orthogonal STBCs. In Fig. \ref{fig_n1}, we present the uncoded BER of 
the LAS detector in decoding $n\times n$ full-rate non-orthogonal STBCs 
from DA in (20.a) for $n=4,8,16$, $\delta=t=1$, and 4-QAM. 
It can be observed that as the STBC code size
$n$ increases, the LAS performs increasingly better such that it achieves
close to SISO AWGN performance (within 0.5 dB at $10^{-3}$ BER and
less) with the $16\times 16$ STBC. We point out that due to the high
complexities involved in decoding large size STBCs using other known
detectors, the BER performance of STBCs with large $n$ has not been
reported in the literature so far. The very fact that we could show
the simulated BER plots (both uncoded as well as turbo coded) for a
$16\times 16$ STBC with 256 complex symbols in one STBC matrix in itself
is a clear indication of the superior low-complexity attribute of the
proposed LAS detector. To our knowledge, we are the first to report the
simulated BER performance of a $16\times 16$ STBC from DA; this became
feasible because of the low-complexity feature of the proposed detector.
In addition, the achievement of near SISO AWGN performance with
$16\times 16$ STBC is a significant result from an implementation
view point as well, since 16 antennas can be easily placed in
communication terminals of moderate size, which can make large MIMO
systems practical.

\subsection{Turbo Coded BER Results for Large STBCs from DA}
\begin{figure}
\centering
\includegraphics[width=3.35in]{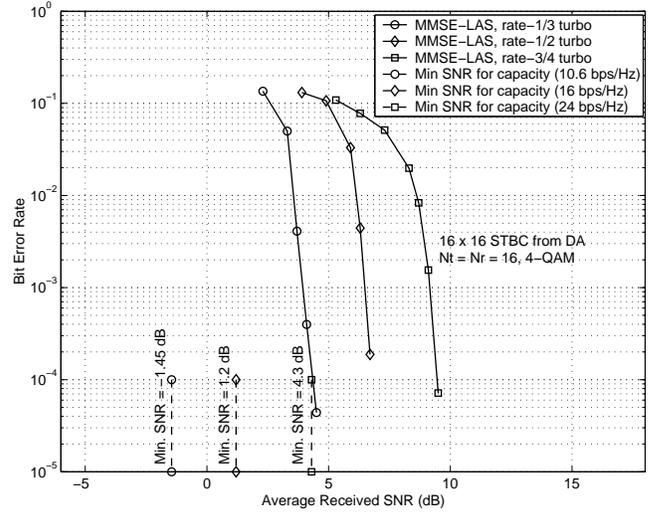}
\caption{Coded BER performance of the proposed LAS detector in decoding
$16 \times 16$ full-rate non-orthogonal STBC from DA. $N_t=N_r=16$. MMSE
initial vector, 4-QAM. Rates of turbo codes: 1/3, 1/2, 3/4.
Proposed LAS detector performs close to within about 5.5 dB from the
theoretical capacity.  } 
\label{fig_n2}
\end{figure}

In Fig. \ref{fig_n2}, we show the coded BER performance of the 
$16\times 16$
STBC using different turbo code rates of 1/3, 1/2, 
\newpage 
and 3/4. With 4-QAM, these turbo code rates along with the 
$16\times 16$ STBC from DA correspond to spectral efficiencies of 10.6 bps/Hz, 
16 bps/Hz and 24 bps/Hz, respectively. The minimum SNRs required to achieve 
these capacities are also shown in Fig. \ref{fig_n2}. It can be observed that 
the proposed detector performs to within about 5.5 dB of the capacity, which 
is an impressive result. In all the turbo coded BER plots in this 
paper, we have used hard decision outputs from the LAS algorithm. In 
\cite{isit08}, we have proposed a method to generate soft decision outputs 
from the LAS algorithm for the individual bits that form the QAM/PAM symbols. 
With the proposed soft decision LAS outputs in \cite{isit08}, the coded
performance is found to move closer to capacity by an additional 1 to
1.5 dB compared to that achieved using hard decision LAS outputs 
reported in this paper.

\section{LAS Detector for Multicarrier CDMA}
\label{sec5}
In this section, we present the proposed LAS detector for multicarrier
CDMA, its performance and complexity. Consider a $K$-user synchronous
multicarrier DS-CDMA system with $M$ subcarriers. Let $b_k \in \{+1,-1\}$
denote the binary data symbol of the $k$th user, which is sent in parallel
on $M$ subcarriers \cite{mccdma1},\cite{mccdma2}.
Let $N$ denote the number of chips-per-bit in the signature waveforms.
It is assumed that the channel is frequency non-selective on each
subcarrier and the fading is slow (assumed constant over one bit interval)
and independent from one subcarrier to the other.

Let ${\bf y}^{(i)}=\left[y_1^{(i)}\,\,y_2^{(i)}\,\,\cdots\,\,y_K^{(i)}\right]^T$denote the $K$-length received signal vector on the $i$th subcarrier; i.e.,
$y_k^{(i)}$ is the output of the $k$th user's matched filter on the
$i$th subcarrier. Assuming that the inter-carrier interference is
negligible, the $K$-length received signal vector on the $i$th
subcarrier ${\bf y}^{(i)}$ can be written in the form
\begin{equation}
\label{eqnA}
{\bf y}^{(i)} \, = \, {{\bf R}^{(i)}{\bf H}^{(i)}{\bf A}{\bf b}}+{\bf n}^{(i)},
\end{equation}
where ${\bf R}^{(i)}$ is the $K\times K$ cross-correlation matrix on
the $i$th subcarrier, 
with its entries $\rho_{lj}^{(i)}$'s denoting the normalized cross 
correlation coefficient
between the signature waveforms of the $l$th and $j$th users on the $i$th
subcarrier. ${\bf H}^{(i)}$ represents the {\normalsize $K\hspace{-0mm}\times
\hspace{-0mm}K$} channel matrix, given by
\begin{eqnarray}
{\bf H}^{(i)} & = & diag\left\{h_1^{(i)}, h_2^{(i)}, \cdots, h_K^{(i)} \right\},\label{eqnC}
\end{eqnarray}
where the channel coefficients $h_k^{(i)}$, $i=1,2,\cdots,M$, are
assumed to be i.i.d. complex Gaussian r.v's (i.e., fade amplitudes are
Rayleigh distributed) with zero mean and
$E\big[\big(h_{kI}^{(i)}\big)^2\big]=E\big[\big(h_{kQ}^{(i)}\big)^2\big]=0.5$,
where $h_{kI}^{(i)}$ and $h_{kQ}^{(i)}$ are the real and imaginary parts
of $h_k^{(i)}$. The $K$-length data vector ${\bf b}$ is given by
\begin{equation}
{\bf b} =  \left[ \begin{array}{cccc}
                b_1 & b_2 & \cdots & b_K
        \end{array} \right]^T,
\end{equation}
and the $K\times K$ diagonal amplitude matrix ${\bf A}$ is given by
\begin{eqnarray}
{\bf A} & = & diag\left\{A_1, A_2, \cdots, A_K \right\},
\end{eqnarray}
where $A_k$ denotes the transmit amplitude of the $k$th user.
The $K$-length noise vector ${\bf n}^{(i)}$ is given by
\begin{equation}
\label{eqnE}
{\bf n}^{(i)} \, = \, \left[\begin{array}{cccc}
n_1^{(i)} & n_2^{(i)} & \cdots & n_K^{(i)} \end{array}\right]^T,
\end{equation}
where $n_k^{(i)}$ denotes the additive noise component of the $k$th user
on the $i$th subcarrier,
which is assumed to be complex Gaussian with zero mean with
{\normalsize $E[n_k^{(i)} \big(n_j^{(i)}\big)^{*}]=\sigma^2$}
when $j=k$ and
{\normalsize $E[n_k^{(i)} \big(n_j^{(i)}\big)^{*}]=\sigma^2\rho_{kj}^{(i)}$}
when $j\neq k$. We assume that all the channel coefficients are perfectly
known at the receiver.

\subsection{LAS Algorithm for MC-CDMA}
\label{sec3aax}
We note that once the likelihood function for the MC-CDMA system in the
above is obtained, it is straightforward to adopt the LAS algorithm
for MC-CDMA. Accordingly, in the multicarrier system considered, the most
likely $\bf b$ is taken as that $\bf b$ which maximizes
\begin{eqnarray}
\Lambda({\bf b}) & = & \sum_{i = 1}^{M}\left({\bf b}^T{\bf A}({\bf H}^{(i)})^*{\bf y}^{(i)} + {\bf b}^T{\bf A}{\bf H}^{(i)}({\bf y}^{(i)})^*\right) \nonumber \\
& & - {\bf b}^T\bigg(\sum_{i = 1}^{M}
{\bf A}{\bf H}^{(i)}{\bf R}^{(i)}({\bf H}^{(i)})^*{\bf A}
\bigg){\bf b}.
\label{lamda1}
\end{eqnarray}
The likelihood function in (\ref{lamda1}) can be written in a form similar
to Eqn. (4.11) in \cite{verdu} as
\begin{eqnarray}
\Lambda({\bf b}) & = & {\bf b}^T{\bf A}{\bf y}_{ceff} - {\bf b}^T{\bf H}_{ceff}{\bf b},
\label{3eq5}
\end{eqnarray}
where
\begin{eqnarray}
{\bf y}_{ceff} &= &\sum_{ i= 1}^{M}\left(\big({\bf H}^{(i)}\big)^*{\bf y}^{(i)} + {\bf H}^{(i)}\big({\bf y}^{(i)}\big)^*\right),\\
\label{3eq5a}
{\bf H}_{ceff} & = & \sum_{ i= 1}^{M}
{\bf A}{\bf H}^{(i)}{\bf R}^{(i)}\big({\bf H}^{(i)}\big)^*{\bf A}.
\label{3eq5b}
\end{eqnarray}
Now observing the similarity between (\ref{3eq5}) and
(\ref{3eq5w}) in Sec. \ref{sec3aa}, the LAS algorithm for MC-CDMA can
be arrived at, along the same lines as that of V-BLAST in the previous
section, with ${\bf y}_{eff}$, ${\bf H}_{eff}$ and ${\bf H}_{real}$ replaced
by ${\bf y}_{ceff}$, ${\bf H}_{ceff}$, and ${\bf H}_{creal}$, respectively,
with all other notations, definitions, and procedures in the algorithm
remaining the same.

\subsection{Complexity of the Proposed Detector for MC-CDMA}
\label{secax}
The complexity of the proposed detector for MC-CDMA can be analyzed
in a similar manner as done for V-BLAST in Sec. \ref{sec2}. First, given
an initial vector, the LAS operation part alone in MC-CDMA has an average
per-bit complexity of $O(MK)$, which is due to $i)$ initial computation
of ${\bf g}(0)$ in (\ref{3eq9w}), which requires $O(MK)$ complexity per
bit, $ii)$ update of ${\bf g}(n)$ in each step as per (\ref{3eq21w}),
which requires $O(K)$ complexity for sequential LAS, and hence constant
per-bit complexity, and $iii)$ the average number of steps required to
reach a fixed point, which, through simulations, is found to have a
constant per-bit complexity. Next, the initial
vector generation using MMSE or ZF has a per-bit complexity of $O(K^2)$
for $K>M$. Finally, combining the above complexities involved in the LAS
part and the initial vector generation part, the overall average per-bit
complexity of the MMSE/ZF-LAS detector for MC-CDMA is $O(K^2)$.
The initial vector generation using MF has a per-bit complexity of
only $O(M)$. Hence, if the MF output is used as the initial vector, then
the overall average per-bit complexity of the MF-LAS is the same as that
of the LAS alone, which is $O(MK)$. For large $K$, the performance of
MF-LAS, ZF-LAS, and MMSE-LAS are almost the same (see Fig. \ref{fig_mc2}),
and hence MF-LAS is preferred because of its linear complexity in number
of users, $K$, for a given $M$.

\subsection{Results and Discussions for MC-CDMA }
\label{lasmcres}
We evaluated the BER performance of the proposed LAS detector for MC-CDMA
through simulations.  We evaluate the uncoded BER performance
of the proposed LAS detector as a function of average SNR, number of users
($K$), number of subcarriers ($M$), and number of chips per bit ($N$). We
also evaluate the BER performance as a function of {\em loading factor},
$\alpha$, where, as done in the CDMA literature \cite{verdu}, we define
$\alpha \Define \frac{K}{MN}$. We call the system as underloaded when
$\alpha < 1$, fully loaded when $\alpha=1$, and overloaded when $\alpha>1$.
Random binary sequences of length $N$ are used as the spreading sequences
on each subcarrier. In order to make a fair comparison between the
performance of MC-CDMA systems with different number of subcarriers, we
keep the system bandwidth the same by keeping $MN$ constant. Also, in that
case we keep the total transmit power to be the same irrespective of the
number of subcarriers used. In the simulation plots we show in this section,
we have assumed that all users transmit with equal amplitude\footnote{We
note that we have simulated the MF/ZF-LAS performance in near-far conditions
as well. Even with near-far effect, the MF/ZF-LAS detectors have been
observed to achieve near single-user performance.}. The LAS algorithm used
is the SLAS with circular checking of bits starting from the first user's
bit.

\begin{figure}
\centering
\includegraphics[width=3.75in]{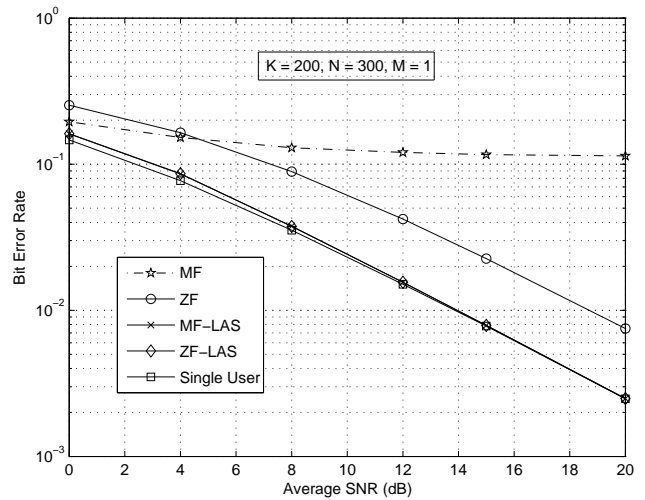}
\caption{BER performance of ZF-LAS and MF-LAS detectors as a function
of average SNR for single carrier CDMA in Rayleigh fading. $M=1$,
$K=200$, $N=300$, i.e., $\alpha=2/3$.}
\label{fig_mc1}
\end{figure}
\begin{figure}
\centering
\includegraphics[width=3.5in]{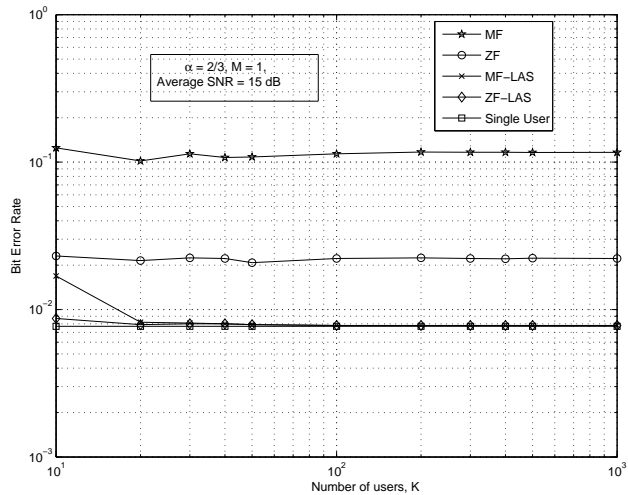}
\caption{BER performance of ZF-LAS and MF-LAS detectors as a function
of number of users, $K$, for single carrier CDMA ($M=1$) in Rayleigh
fading for a fixed $\alpha=2/3$ and average SNR = 15 dB. $N$ varied
from 15 to 1500.}
\label{fig_mc2}
\end{figure}

First, in Fig. \ref{fig_mc1}, we present the BER performance of
MF/ZF-LAS detectors as a function of average SNR
in a single carrier (i.e., $M=1$) {\em underloaded system}, where we
consider $\alpha=2/3$ by taking $K=200$ users and $N=300$ chips per bit.
For comparison purposes, we also plot the performance of MF and ZF without
LAS. Single user (SU) performance, which corresponds to the case of no
multiuser interference (i.e., $K=1$), is also shown as a lower bound on
the achievable multiuser performance. From Fig. \ref{fig_mc1}, we can
observe that the performance of MF and ZF detectors are far away from
the SU performance. Whereas, the ZF-LAS as well as MF-LAS detectors almost
achieve the SU performance. We point out that, like ZF detector, other
suboptimum detectors including MMSE, SIC, and PIC detectors \cite{verdu}
also do not achieve near SU performance for the considered loading factor
of 2/3, whereas the MF-LAS detector achieves near SU performance, that too
at a lesser complexity than these other suboptimum detectors.

Next, in Fig. \ref{fig_mc2}, we show the BER performance of the MF/ZF-LAS
detectors for $M=1$ as a function of number of users, $K$, for a fixed
value of $\alpha=2/3$ at an average SNR of 15 dB. We varied $K$ from 10 to
1000 users.  SU performance is also shown (as the bottom most horizontal
line) for comparison. It can be seen that, for the fixed value of
$\alpha=2/3$, both the MF-LAS as well as the ZF-LAS achieve near SU
performance (even in the presence of 1000 users), whereas the ZF and MF
detectors do not achieve the SU performance.

\begin{figure}
%\centering
\hspace{-3mm}
\includegraphics[width=3.5in]{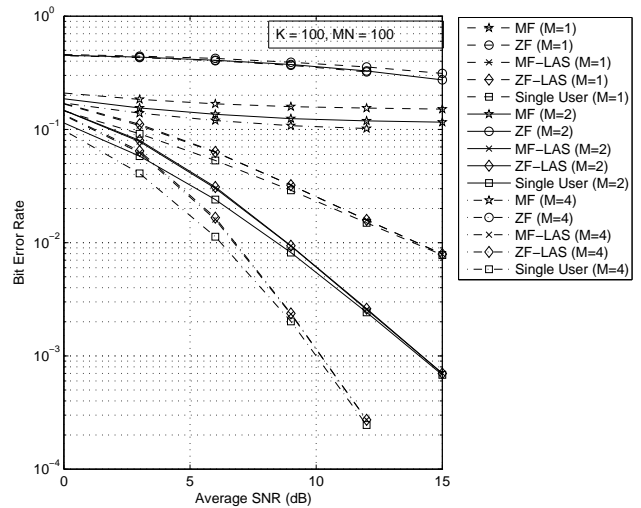}
\caption{BER performance of ZF-LAS and MF-LAS detectors as a function
of average SNR for multicarrier CDMA in Rayleigh fading. $M=1,2,4$,
$\alpha=1$, $K=100$, $MN=100$.}
\label{fig_mc3}
\end{figure}

In Fig. \ref{fig_mc3}, we show the BER performance of the MF/ZF-LAS
detectors as a function of average SNR for different number of subcarriers,
namely, $M=1,2,4$, keeping a constant $MN=100$, for a {\em fully loaded
system} (i.e., $\alpha=1$) with $K=100$. Keeping $\alpha=1$ and $K=100$
for all cases means that $i)$ $N=100$ for $M=1$, $ii)$ $N=50$ for $M=2$,
and $iii)$ $N=25$ for $M=4$. The SU performance for $M=1$ (1st order
diversity), $M=2$ (2nd order diversity), and $M=4$ (4th order diversity)
are also plotted for comparison. These diversities are essentially due to
the frequency diversity effect resulting from multicarrier combining of
signals from $M$ subcarriers. It is interesting to see that even in a
fully loaded system, the MF/ZF-LAS detectors achieve all the frequency
diversity possible in the system (i.e., MF/ZF-LAS detectors achieve SU
performance with 1st, 2nd and 4th order diversities for $M=1,2$ and 4,
respectively). On the other hand, ZF detector is unable to achieve the
frequency diversity in the fully loaded system, and its performance is
very poor compared to MF/ZF-LAS detectors.

\begin{figure}
\centering
\includegraphics[width=3.75in]{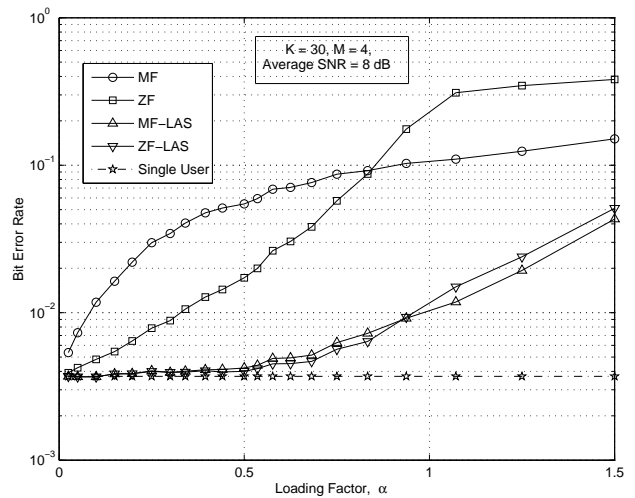}
\caption{BER performance of ZF-LAS and MF-LAS detectors as a function
of loading factor, $\alpha$, for multicarrier CDMA in Rayleigh
fading. $M=4$, $K=30$, $N$ varied from 300 to 5, average SNR = 8 dB. }
\label{fig_mc4}
\end{figure}

Next, in Fig. \ref{fig_mc4}, we present the BER performance of ZF/MF-LAS
detectors in a MC-CDMA system with $M=4$ as a function of loading factor,
$\alpha$, where we vary $\alpha$ from $0.025$ to 1.5. We realize this
variation in $\alpha$ by fixing $K=30$, $M=4$, and varying $N$ from 300
to 5. The average SNR considered is 8 dB. From Fig. \ref{fig_mc4}, it
can be observed that as $\alpha$ increases all detectors loose performance,
but the MF/ZF-LAS detectors can offer relatively good performance even at
{\em overloaded conditions} of $\alpha>1$. Another observation is that at
$\alpha>1$, MF-LAS performs slightly better than ZF-LAS. This is because
$\alpha>1$ corresponds to a high interference condition, and it is known
in MUD literature \cite{verdu} that ZF can perform worse than MF at
low SNRs and high interference. In such cases, starting with a better
performing MF output as the initial vector, MF-LAS performs better.

Further to our present work on the application of MF/ZF-LAS detectors for
MC-CDMA, several extensions are possible on the practical application of
these detectors in CDMA systems. Two such useful extensions are $i)$
MF/ZF/MMSE-LAS for frequency selective CDMA channels with RAKE combining;
we point out that a similar approach and system model adopted here for
MC-CDMA is applicable, by taking a view of equivalence between frequency
diversity through MC combining and multipath diversity through RAKE
combining, and $ii)$ MF/ZF/MMSE-LAS for asynchronous CDMA systems, which
can be carried out once the system model is appropriately written
in a form similar to (\ref{eqnA}). These two extensions can allow
MF/ZF-LAS detectors to be practical in CDMA systems (e.g., 2G and 3G
CDMA systems), with potential for significant gains in system capacity.
Current approaches to MUD considered in practical CDMA systems appear
to be mainly PIC and SIC. However, the illustrated fact that MF-LAS can 
easily outperform PIC/SIC detectors both in performance and complexity 
for large number of users suggests that MF-LAS can be a powerful MUD 
approach in practical CDMA systems.

\section{Conclusions}
\label{sec6}
We presented a near-capacity achieving, low-complexity detector
for large MIMO systems having tens to hundreds of antennas, and showed
its uncoded/coded BER performance in the detection of V-BLAST and in the
decoding of full-rate non-orthogonal STBCs from DA. The proposed detector 
was shown to have excellent attributes in terms of both low complexity as
well as nearness to theoretical capacity performance, achieving high spectral 
efficiencies of the order of tens to hundreds of bps/Hz. To our knowledge, 
our reporting of the decoding of a large full-rate non-orthogonal STBC like 
$16\times 16$ STBC from DA and its BER/nearness to capacity results is for 
the first time in the literature. We further point out that the proposed 
detector has good potential for application in practical MIMO wireless 
standards, e.g., the low-complexity feature of the proposed detector can 
allow the inclusion of $4\times 4$, $8\times 8$, $16\times 16$ 
non-orthogonal STBCs from DA into MIMO wireless standards like IEEE 
802.11n and IEEE 802.16e, which, in turn, can achieve higher spectral 
efficiencies than those are currently possible in these standards.

We conclude this paper by pointing to the following remark made by the
author of \cite{jafarkhani} in its preface in 2005: {\em ``It was just a
few years ago, when I started working at AT\&T Labs -- Research, that
many would ask `who would use more than one antenna in a real system?'
Today, such skepticism is gone.''} Extending this sentiment, we believe
large MIMO systems would be practical in the future, and the practical
feasibility of low-complexity detectors like the one we presented in
this paper could be a potential trigger to create wide interest in the
implementation of large MIMO systems. For example, antenna/RF technologies 
and channel estimation for large MIMO systems could open up as new focus 
areas. Potential large MIMO applications include inter-base station/base
station controller back-haul connectivity using large MIMO links, and 
wireless IPTV. Other interesting large MIMO applications can be thought 
of as well.

\begin{biographynophoto}
%[{\includegraphics[width=1in,height=1.25in,clip,keepaspectratio]{vishnu_vardhan_photo}}]
{K. Vishnu Vardhan}
was born in Andhra Pradesh, India. He received the undergraduate degree 
in Electronics and Communication Engineering from Pondicherry University, 
Pondicherry, India, in 2005. He received the postgraduate degree in 
Telecommunication Engineering from Indian Institute of Science, Bangalore, 
India, in 2007. Since July 2007, he has been with Cisco Systems (India) 
Private Limited, Bangalore, India. His research interests include 
multiuser detection and low-complexity detectors for CDMA and MIMO systems.
\end{biographynophoto}

%\vspace{-5mm}
\begin{biographynophoto}
%[{\includegraphics[width=1in,height=1.25in,clip,keepaspectratio]{saif_k_mohammed_photo}}]
{Saif Khan Mohammed} 
received his B.Tech degree in Computer Science and
Engineering from the Indian Institute of Technology, New Delhi, India,
in 1998. From 1998 to 2000, he was employed with Philips Inc., Bangalore,
India, as an ASIC design engineer. From 2000 to 2003, he worked with Ishoni
Networks Inc., Santa Clara, CA, as a senior chip architecture engineer.
From 2003 to 2007, he was employed with Texas Instruments, Bangalore as
systems and algorithms designer in the wireless systems group. He is 
currently pursuing his doctoral degree in Electrical and Communication
Engineering at the Indian Institute of Science, Bangalore, India. His
research interests include low-complexity detection, estimation and
coding for wireless communications systems.
\end{biographynophoto}

\vfill
\begin{biographynophoto}
%[{\includegraphics[width=1in,height=1.25in,clip,keepaspectratio]{chockalingam_photo}}]
{A. Chockalingam} was born in Rajapalayam, Tamil Nadu, India. He
received the B.E. (Honors) degree in Electronics and Communication
Engineering from the P. S. G. College of Technology, Coimbatore,
India, in 1984, the M.Tech degree with specialization in satellite
communications from the Indian Institute of Technology, Kharagpur,
India, in 1985, and the Ph.D. degree in Electrical Communication
Engineering (ECE) from the Indian Institute of Science (IISc),
Bangalore, India, in 1993. During 1986 to 1993, he worked with the
Transmission R \& D division of the Indian Telephone Industries
Limited, Bangalore. From December 1993 to May 1996, he was a
Postdoctoral Fellow and an Assistant Project Scientist at the
Department of Electrical and Computer Engineering, University of
California, San Diego. From May 1996 to December 1998, he served
Qualcomm, Inc., San Diego, CA, as a Staff Engineer/Manager in the
systems engineering group. In December 1998, he joined the faculty
of the Department of ECE, IISc, Bangalore, India, where he is an
Associate Professor, working in the area of wireless communications
and networking.
Dr. Chockalingam is a recipient of the Swarnajayanti Fellowship from
the Department of Science and Technology, Government of India. He
served as an Associate Editor of the IEEE Transactions on Vehicular
Technology from May 2003 to April 2007. He currently serves as an
Editor of the IEEE Transactions on Wireless Communications. He is a
Fellow of the Indian National Academy of Engineering.
\end{biographynophoto}

\vspace{-85mm}
\begin{biographynophoto}
%[{\includegraphics[width=1in,height=1.25in,clip,keepaspectratio]{sundar_rajan_photo}}]
{B. Sundar Rajan} was born in Tamil Nadu, India.
He received the B.Sc. degree in mathematics from Madras University,
Madras, India, the B.Tech degree in electronics from Madras Institute
of Technology, Madras, and the M.Tech and Ph.D. degrees in electrical
engineering from the Indian Institute of Technology, Kanpur, India,
in 1979, 1982, 1984, and 1989 respectively. He was a faculty member
with the Department of Electrical Engineering at the Indian Institute
of Technology in Delhi, India, from 1990 to 1997. Since 1998, he has
been a Professor in the Department of Electrical Communication
Engineering at the Indian Institute of Science, Bangalore, India.
His primary research interests include space-time coding for MIMO
channels, distributed space-time coding and cooperative communication,
coding for multiple-access and relay channels, with emphasis on
algebraic techniques.

Dr. Rajan is an Associate Editor of the IEEE Transactions on
Information Theory, an Editor of the IEEE Transactions on Wireless
Communications, and an Editorial Board Member of International Journal
of Information and Coding Theory. He served as Technical Program
Co-Chair of the IEEE Information Theory Workshop (ITW'02), held in
Bangalore, in 2002. He is a Fellow of the Indian National Academy of
Engineering and recipient of the IETE Pune Center's S.V.C Aiya Award
for Telecom Education in 2004. Dr. Rajan is a Member of the American
Mathematical Society.
\end{biographynophoto}

\end{document}